\documentclass[aps,prl,twocolumn,showpacs,preprintnumbers,superscriptaddress]{revtex4}
\usepackage{amssymb}
\usepackage{graphicx}
\usepackage{dcolumn}
\usepackage{amsmath}
\usepackage{color}
\usepackage{bm}

\begin{document}

\title{Quantum Superconductor-Metal Transition in Al, C doped MgB$_{2}$ and
overdoped Cuprates?}
\author{T. Schneider}
\affiliation{Physik-Institut der Universit\"{a}t Z\"{u}rich,
Winterthurerstrasse 190, CH-8057 Z\"urich, Switzerland}

\begin{abstract}
We consider the realistic case of a superconductor with a nonzero
density of elastic scatterers, so that the normal state conductivity
is finite. The quantum superconductor-metal (QSM) transition can
then be tuned by varying either the attractive electron-electron
interaction, the quenched disorder, or the applied magnetic field.
We explore the consistency of the associated scaling relations,
$T_{c}\propto \lambda \left( 0\right) ^{-1}\propto \Delta \left(
0\right) \propto \xi \left( 0\right) ^{-1}\propto H_{c2}\left(
0\right) ^{1/2}$ and $T_{c}\left( H\right) \propto \lambda \left(
0,\text{ }H\right) ^{-1}\propto \Delta \left( 0,\text{ }H\right)
\propto \left( H_{c2}\left( 0\right) -H\right) ^{1/2}$, valid for
all dimensions $D>2$, with experimental data, in Al, C doped
MgB$_{2}$ and overdoped cuprates.

\end{abstract}

%\pacs{74.70.Ad, 74.25.Op, 74.25.Ha, 76.75.+i, 83.80.Fg}
\maketitle

%\preprint{PREPRINT (\today)}

%%
%

%%
%

%\section{Introduction}

Understanding the phenomenon of superconductivity, now observed in
quite disparate systems, such as simple elements, fullerenes,
molecular metals, cuprates, borides, \textit{etc.}, involves
searching for universal relations between superconducting properties
across different materials, which might provide hints towards a
unique classification. In spite of the great impact of the BCS
theory \cite{bcs}, the discovery of superconductivity in the
cuprates in 1986 \cite{bednorz} made it clear that the BCS relations
between the critical amplitudes of the gap ($\Delta _{0}$ ), the
correlation length ($\xi _{0}$), the magnetic penetration depth
($\lambda _{0}$), the upper critical field ($H_{c20}$) and the
transition temperature $T_{c}$, namely \cite{ketterson}
\begin{equation}
T_{c}\propto \lambda _{0}^{-1}\propto \Delta _{0}^{-1}\propto \xi
_{0}^{-1}\propto H_{c20}^{1/2},  \label{eq1}
\end{equation}
Here, $\lambda \left( T\right) =\lambda _{0}t^{1/2}$, $\Delta \left(
T\right) =\Delta _{0}t^{1/2}\simeq 1.76\Delta \left( 0\right)
t^{1/2}$, $\xi \left( T\right) =\xi _{0}t^{-1/2}$, and
$H_{c2}=H_{c20}t$, close to the superconductor metal transition,
with $t=1-T/T_{c}$ and $2\Delta \left( 0\right) /\left(
k_{B}T_{c}\right) \simeq 3.52$. Furthermore, there are empirical
relations between $T_{c}$ and the zero-temperature superfluid
density, $\rho _{s}(0)$, related to the zero-temperature magnetic
field penetration depth $\lambda \left( 0\right) $ in terms of $\rho
_{s}(0)\propto \lambda ^{-2}(0)$. In various families of underdoped
cuprate superconductors there is the empirical relation
$T_{c}\propto \rho _{s}(0)\propto \lambda ^{-2}(0)$, first
identified by Uemura \textit{et al.} \cite{Uemura89,Uemura91}, while
in molecular superconductors, $T_{c}\propto \lambda ^{-3}(0)$,
appears to apply \cite{pratt}. Both scaling forms appear to have no
counterpart in the BCS scenario and even in more conventional
superconductors, including Mg$_{1-x}$Al$_{x}$B$_{2}$,
Mg(C$_{x}$B$_{1-x}$)$_{2}$, and MgB$_{2+x}$, such relationships
remain to be explored.

According to the theory of quantum critical phenomena a power law
relation between $T_{c}$ and $\rho _{s}(0)\propto \lambda ^{-2}(0)$
is expected whenever there is a critical line $T_{c}\left( x\right)
$ with a critical endpoint $x=x_{c}$ \cite{kim,tsbook,parks}. Here
the transition temperature vanishes and a quantum phase transition
occurs. $x$ denotes the tuning parameter of the transition. A
variety of underdoped cuprate superconductors exhibits such a
critical line, ending at the quantum superconductor to insulator
(QSI) transition, where the materials become essentially two
dimensional \cite{tsphysb}. If the finite temperature behavior in
this regime is controlled by the $3D-xy$ critical point, $T_{c}$ and
$\rho _{s}(0)$ scale as \cite{kim,tsbook,parks}
\begin{equation}
T_{c}\propto \rho _{s}(0)^{z/(D+z-2)}.  \label{eq2}
\end{equation}
$z$ denotes the dynamic critical exponent of the quantum transition
in $D$ dimensions. For $D=2$ we recover the empirical Uemura
relation, $T_{c}\propto \rho _{s}(0)$ \cite{Uemura89,Uemura91},
irrespective of the value of $z$.

\begin{figure}[htb]
\includegraphics[width=0.9\linewidth]{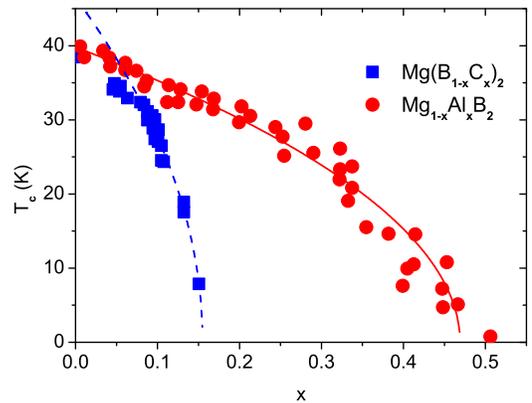}
\caption{(Color online) $T_{c}$ \textit{vs}. the nominal
concentration $x$ for Mg$_{1-x}$Al$_{x}$B$_{2}$ and
Mg(C$_{x}$B$_{1-x}$)$_{2}$ taken from Postorino \textit{et
al}.\cite{postorino} and Gonnelli \textit{et al}.\cite{gonnelli}.}
\label{fig1}
\end{figure}

There is also considerable evidence for a critical line $T_{c}\left(
x\right) $ in more conventional superconductors, including
Mg$_{1-x}$Al$_{x}$B$_{2}$, Mg(C$_{x}$B$_{1-x}$)$_{2}$,
MgB$_{2-x}$Be$_{x}$, and NbB$_{2+x}$, where superconductivity
disappears at some critical value
$x=x_{c}$\cite{postorino,gonnelli,ahn,Escamilla06}, whereupon a
quantum superconductor to metal (QSM) transition is expected to
occur. To illustrate this behavior we depicted in Fig. \ref{fig1}
the data for $T_{c}$ \textit{vs}. the nominal concentration $x$ for
Mg$_{1-x}$Al$_{x}$B$_{2}$ and Mg(C$_{x}$B$_{1-x}$)$_{2} $ taken from
Postorino \textit{et al}.\cite{postorino} and Gonnelli \textit{et
al}.\cite{gonnelli}.

Since in these nearly isotropic materials the anisotropy does not
change substantially upon substitution the QSM transition occurs in
$D=3$. Furthermore, there is considerable evidence that on
increasing the Al or C content the homogeneity and the
crystallographic order decrease even in segregation-free
samples\cite{connelliprb,klein,daghero}.

Since in these nearly isotropic materials the anisotropy does not
change substantially upon substitution the QSM transition occurs in
$D=3$. Furthermore, there is considerable evidence that on
increasing the Al or C content the homogeneity and the
crystallographic order decrease even in segregation-free
samples\cite{connelliprb,klein,daghero}.

For this reason we consider the realistic case of a superconductor
with a nonzero density of elastic scatterers, so that the normal
state conductivity is finite. In the absence of an applied magnetic
field the QSM transition can then be tuned by varying either the
attractive electron-electron interaction, or the quenched disorder.
In a theoretical description, quenched disorder can occur on a
microscopic level, e.g., due to randomly distributed scattering
centers. For such systems it was shown that the upper critical
dimension $D_{c}^{+}$, above which the critical behavior is governed
by a simple Gaussian fixed point (FP), is lower than that of the
corresponding classical or finite temperature transition,namely
$D_{c}^{+}=2$ \cite{kirkpatrick,zhou}. For $D>2$ the transition is
then governed by a Gaussian FP with unusual properties. Since the
mean-field/Gaussian theory yields the exact \ critical behavior at
$T=0$, all relations between observables that are derived at finite
temperature within BCS theory are valid. Accordingly, the zero
temperature counterpart of the scaling relation reads
\begin{equation}
T_{c}\propto \lambda \left( 0\right) ^{-1}\propto \Delta \left(
0\right) \propto \xi \left( 0\right) ^{-1}\propto H_{c2}\left(
0\right) ^{1/2}, \label{eq3}
\end{equation}
while $T_{c}$ and the dimensionless distance from the critical point
$\delta $ ($\delta <0$ in the disordered phase) are related by
\cite{kirkpatrick}
\begin{equation}
T_{c}\propto \exp (-1/\left\vert \delta \right\vert ).  \label{eq4}
\end{equation}%
Hyperscaling is violated by the usual mechanism that is operative
above an upper critical dimension. Indeed, this QSM transition
occurs in $D=3$, while the upper critical dimension of the QSM
transition is $D_{c}^{+}=2$ \cite{kirkpatrick,zhou}. For this reason
the scaling relation (\ref{eq2}), involving hyperscaling, does not
apply in the present case where $z=2$ \cite{kirkpatrick}.

\begin{figure}[htb]
\includegraphics[width=0.9\linewidth]{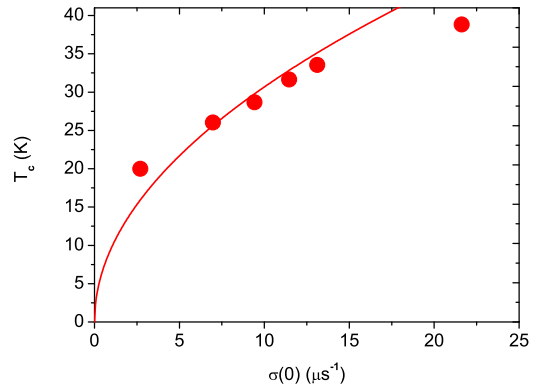}
\caption{(Color online) $T_{c}$ \textit{vs}. zero-temperature
muon-spin depolarization rate $\sigma (0)\propto \rho _{s}(0)\propto
\lambda ^{-2}(0)$ for Mg$_{1-x}$Al$_{x} $B$_{2}$ derived from
Serventi \textit{et al}.\cite{Serventi04}. The solid line is
Eq.(\ref{eq3}) in terms of $T_{c}=9.7\sigma _{sc}(0)^{1/2}$.}
\label{fig2}
\end{figure}

We are now prepared to confront the scaling predictions for a
disorder tuned QSM transition with experiment. In Fig. \ref{fig2} we
show $T_{c}$ \textit{vs}. zero-temperature muon-spin depolarization
rate $\sigma (0)\propto \rho _{s}(0)\propto \lambda ^{-2}(0)$ for
Mg$_{1-x}$Al$_{x}$B$_{2}$ taken from Serventi \textit{et al}.
\cite{Serventi04}. Although the data is rather sparse, in particular
close to the QSM transition, the reduction of the superfluid density
$\sigma (0)\propto \rho _{s}(0)\propto \lambda ^{-2}(0)$ with
decreasing $T_{c}$ is clearly observed, consistent with the flow to
the QSM transition where $T_{c}$ and $\sigma \left( 0\right) $ scale
as $T_{c}\propto 1/\lambda \left( 0\right) \propto \sigma (0)^{1/2}$
(Eq.(\ref{eq3})).

\begin{figure}[htb]
%\centering
\includegraphics[width=0.9\linewidth]{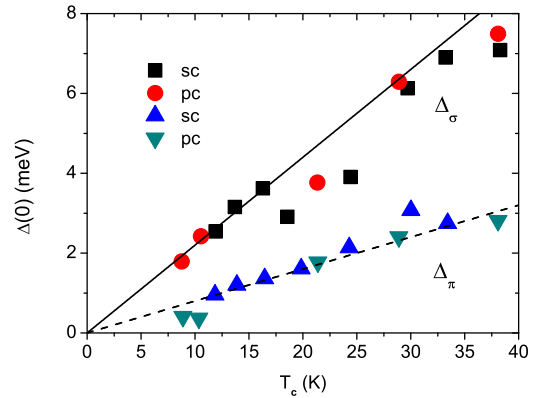}
%\includegraphics[width=0.7\linewidth]{fig3}
% \vspace{-1.0cm}
\caption{(Color online) $\Delta _{\sigma }\left( 0\right) $ and
$\Delta _{\pi }\left( 0\right) $ \textit{vs}. $T_{c}$ for
Mg$_{1-x}$Al$_{x}$B$_{2}$ single crystals (sc) and polycrystals (pc)
taken from Daghero \textit{et al}.\cite{daghero}. The solid and
dashed lines are Eq.(\ref{eq3}) in terms of  $\Delta _{\sigma
}\left( 0\right) =0.22T_{c}$ and $\Delta _{\sigma }\left( 0\right)
=0.08T_{c}$.} \label{fig3}
\end{figure}

To substantiate this supposition further, we consider the $T_{c}$
dependence of the gap $\Delta \left( 0\right) $. In Fig.\ref{fig3}
we depicted the experimental data of Daghero \textit{et
al}.\cite{daghero} for the two gap superconductor
Mg$_{1-x}$Al$_{x}$B$_{2}$. Although the data does not extend very
close to the QSM transition, the flow to $T_{c}\propto \Delta \left(
0\right) $ (Eq.(\ref{eq3})) can be anticipated.

\begin{figure}[htb]
\includegraphics[width=0.9\linewidth]{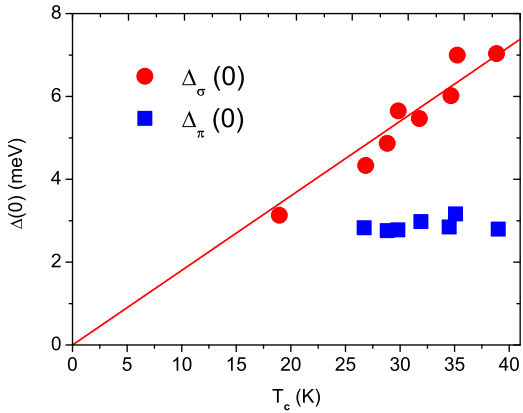}
\caption{(Color online) $\Delta _{\sigma }\left( 0\right) $ and
$\Delta _{\pi }\left( 0\right) $ \textit{vs}. $T_{c}$ for
Mg(B$_{1-x}$C$_{x}$)$_{2}$ single crystals taken from Gonnelli
\textit{et al}.\cite{connelliprb}. The solid line is Eq.(\ref{eq3})
in terms of $\Delta _{\sigma }\left( 0\right) =0.18T_{c}$.}
\label{fig4}
\end{figure}

According to Fig.\ref{fig4}, showing $\Delta _{\sigma }\left(
0\right) $ and $\Delta _{\pi }\left( 0\right) $ \textit{vs}. $T_{c}$
for Mg(B$_{1-x}$C$_{x}$)$_{2}$ single crystals taken from Gonnelli
\textit{et al}.\cite{connelliprb}, the flow to the QSM transition
(Eq.(\ref{eq3})) is apparent in the $\sigma $-gap as well, while the
$\pi $-gap, nearly constant down to $T_{c}=19$ K, appears to merge
the $\sigma $-gap below this transition
temperature\cite{connelliprb}.

\begin{figure}[htb]
\includegraphics[width=0.9\linewidth]{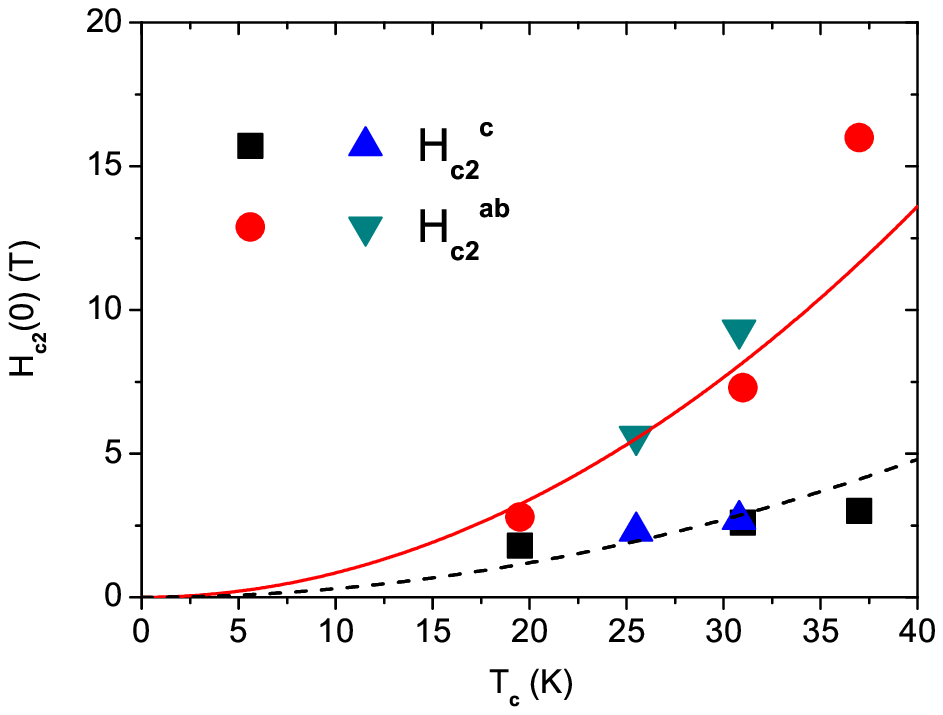}
\caption{(Color online) $H_{c2}^{ab}\left( 0\right) $ and
$H_{c2}^{c}\left( 0\right) $ \textit{vs}. $T_{c}$ for
Mg$_{1-x}$Al$_{x}$B$_{2}$ single crystals taken from Klein
\textit{et al}. $\left( \blacksquare ,\bullet \right) $ \cite{klein}
and Kim \textit{et al}. $\left( \blacktriangle ,\blacktriangledown
\right) $ \cite{kim}. The solid and dashed lines are Eq.(\ref{eq3})
in terms of $H_{c2}^{ab}\left( 0\right) =0.0085T_{c}^{2}$ and
$H_{c2}^{c}\left( 0\right) =0.003T_{c}^{2}$.} \label{fig5}
\end{figure}

Next we turn to the $T_{c}$ dependence of the upper critical fields
$H_{c2}^{ab}\left( 0\right) $ and $H_{c2}^{c}\left( 0\right) $.
Although the experimental data for Mg$_{1-x}$Al$_{x}$B$_{2}$ single
crystals taken from Klein \textit{et al}. \cite{klein} and Kim
\textit{et al}. \cite{heon} shown in Fig. \ref{fig5} is rather
sparse, consistency towards QSM scaling behavior
$H_{c2}^{ab,c}\left( 0\right) \propto T_{c}^{2}$ (Eq.(\ref{eq3})),
indicated by the solid and dashed lines, can be anticipated. From
these lines we deduce for the zero temperature anisotropy the
estimate $\gamma \left( 0\right) =\left( H_{c2}^{c}\left( 0\right)
/H_{c2}^{ab}\left( 0\right) \right) ^{1/2}=\xi _{ab}\left( 0\right)
/\xi _{c}\left( 0\right) \simeq 1.9$. Because $H_{c2}^{ab}\left(
0\right) \propto \xi _{c}\left( 0\right) ^{-2}\propto T_{c}^{2}$ and
$H_{c2}^{c}\left( 0\right) \propto \xi _{ab}\left( 0\right)
^{-2}\propto T_{c}^{2}$ the reduction of the upper critical fields
mirrors the increase of the correlation lengths as the QSM
transition is approached.

\begin{figure}[htb]
\includegraphics[width=0.9\linewidth]{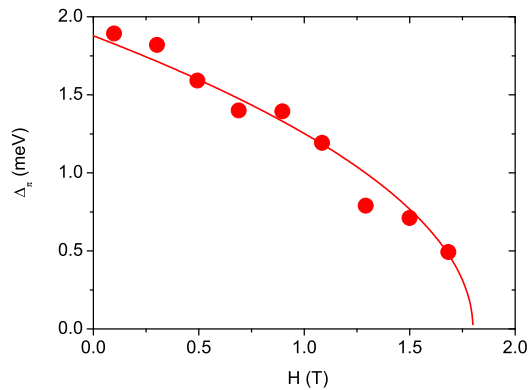}
\caption{(Color online) $\Delta _{\pi }\left( T=6.5\text{ K,
}x=0.2\right) $ \textit{vs}. $H$ applied along the $c$-axis of
Mg$_{1-x}$Al$_{x}$B$_{2}$ single crystals taken from Giubileo
\textit{et al}. \cite{giubileo}. The solid line is Eq. (\ref{eq5})
in terms of $\Delta _{\pi }\left( T=6.5\text{ K, }x=0.2\right)
=1.4\left( H_{c2}^{c}\left( 0,\text{ }x=0.2\right) -H\right) ^{1/2}$
with $H_{c2}^{c}\left( 0,\text{ }x=0.2\right) =1.8$ T.} \label{fig6}
\end{figure}

Finally we consider the magnetic field tuned QSM transition for
fixed $x$. Noting that the correlation length and the magnetic field
scale as $\xi \left( 0\right) ^{2}\left( H_{c2}\left( 0,x\right)
-H\right) \propto \Phi _{0}$, together with Eq. (\ref{eq4}), the
zero temperature gap scales then close to the QSM transition as
\begin{equation}
\Delta \left( 0,\text{ }x\right) \propto \left( H_{c2}\left(
0,\text{ } x\right) -H\right) ^{1/2}.  \label{eq5}
\end{equation}
Here the critical line $T_{c}\left( x,H\right) $ ends at
$H_{c2}\left( T=0,x\right) $. In Fig.\ref{fig6} we depicted $\Delta
_{\pi }\left( T=6.5 \text{ K}\right) $ \textit{vs}. $H$ applied
along the $c$-axis of a Mg$_{1-x} $Al$_{x}$B$_{2}$ single crystals
with $x=0.2$ ($T_{c}\simeq 24$ K) taken from Giubileo \textit{et
al}. \cite{giubileo}. Although Eq. (\ref{eq5}) represents the
asymptotic behavior we observe remarkable agreement with the local
tunneling data over the entire magnetic field range. Clearly, the
occurrence of the magnetic field tuned QSM transition is not
restricted to the gap. From Eqs. (\ref{eq3}) and (\ref{eq5}) we
deduce
\begin{eqnarray}
T_{c}\left( x,\text{ }H\right)  &\propto &\lambda \left( 0,\text{
}x,\text{ }H\right) ^{-1}\propto \Delta \left( 0,\text{ }x,\text{
}H\right)   \nonumber\\&\propto &\left( H_{c2}\left( 0,\text{
}x\right) -H\right) ^{1/2}, \label{eq6}
\end{eqnarray}
which remains to be tested experimentally.

\begin{figure}[htb]
\includegraphics[width=0.9\linewidth]{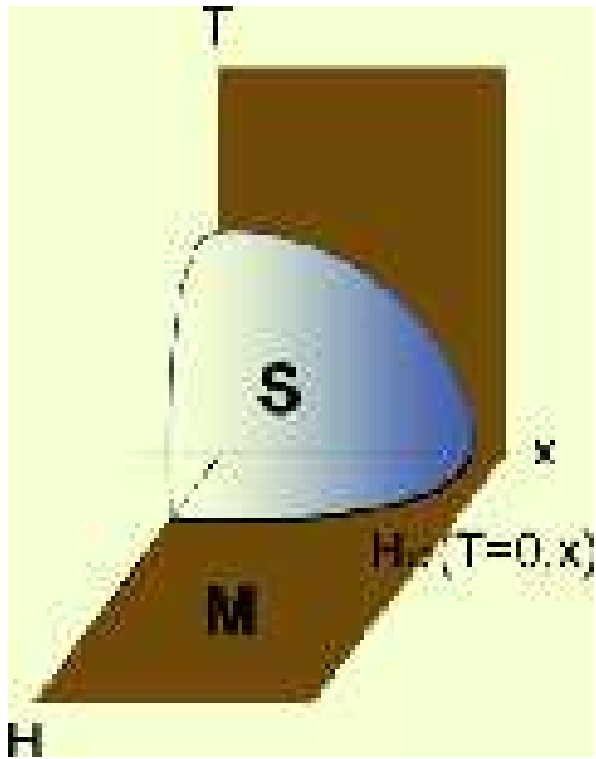}
\caption{(Color online) Schematic phase diagram. There is the
surface of finite temperature superconductor (S) to metal (M)
transitions ending at the critical line $H_{c2}(T=0,x)$ of quantum
superconductor-metal (QSM) transitions.} \label{fig7}
\end{figure}

The corresponding schematic phase diagram is shown in
Fig.\ref{fig7}. As the substituent concentration or the magnetic
field is increased $T_{c}$ is suppressed and driven all the way to
zero, where along the line $H_{c2}\left( T=0,x\right) $ the QSM
transition, characterized by the scaling relations (\ref{eq3}) and
(\ref{eq6}) occurs.

These scaling relations also imply that close to the QSM transition
the isotope and pressure effects on these observables are not
independent of one another. From Eqs. (\ref{eq3}) and (\ref{eq6}) we
deduce for the relative changes upon isotope substitution or applied
pressure the relations
\begin{equation}
\frac{\Delta T_{c}}{T_{c}}=\frac{\Delta a}{a}-\frac{\Delta \lambda
\left( 0\right) }{\lambda \left( 0\right) }=\frac{\Delta
b}{b}+\frac{\Delta \Delta \left( 0\right) }{\Delta \left( 0\right)
}=\frac{\Delta c}{c}+\frac{\Delta H_{c2}\left( 0\right)
}{2H_{c2}\left( 0\right) },  \label{eq7}
\end{equation}%
where $T_{c}=a/\lambda \left( 0\right) =b\Delta \left( 0\right)
=cH_{c2}\left( 0\right) ^{1/2}$ and
\begin{eqnarray}
\frac{\Delta T_{c}\left( x,\text{ }H\right) }{T_{c}\left( x,\text{
}H\right) } &=&\frac{\Delta d}{d}-\frac{\Delta \lambda \left(
0,\text{ }x,\text{ }H\right) }{\lambda \left( 0,\text{ }x,\text{
}H\right) }=\frac{\Delta e}{e}+\frac{\Delta \Delta \left( 0,\text{
}x,\text{ }H\right) }{\Delta \left( 0,\text{ }x,\text{ }H\right) }  \nonumber \\
&=&\frac{\Delta f}{f}+\frac{\Delta H_{c2}\left( 0,\text{ }x\right)
}{2H_{c2}\left( 0,\text{ }x\right) },  \label{eq8}
\end{eqnarray}
where $T_{c}\left( x,\text{ }H\right) =d/\lambda \left( 0\right)
=e\Delta \left( 0,\text{ }x,\text{ }H\right) =fH_{c2}\left( 0,\text{
}x\right) ^{1/2}$. $a$ to $f$ are non-universal coefficients.

We sketched, following Kirkpatrick and Belitz \cite{kirkpatrick} the
scaling relations of a quantum superconductor to metal (QSM)
transition for nearly isotropic three dimensional systems,
considering the realistic case of a superconductor with a nonzero
density of elastic scatterers, so that the normal state conductivity
is finite. The QSM transition can then be tuned by varying either
the attractive electron-electron interaction, the quenched disorder,
or the applied magnetic field. We have shown that Mg$_{1-x}$Al$_{x}
$B$_{2}$ and Mg(B$_{1-x}$C$_{x}$)$_{2}$, where increasing Al or C
enhances
the disorder even in segregation-free samples \cite%
{connelliprb,klein,daghero}, are potential candidates to observe
this QSM transition, characterized by the scaling relations
(\ref{eq3}), (\ref{eq6}), (\ref{eq7}), and (\ref{eq8}). Indeed, as a
whole, the spare experimental data points to this QSM transition,
but more extended experimental data are needed to confirm this
characteristic scaling relation unambiguously. Indeed, based on band
structure calculations and the Eliashberg theory, it was argued that
the observed decrease of $T_{c}$ of Al and C doped MgB$_{2}$ samples
can be understood mainly in terms of a band filling effect due to
the electron doping by Al and C \cite{pena,kortus}. Finally we note
that NbB$_{2+x}$ \cite{Escamilla06,Takagiwa04,khasanov},
Nb$_{1-x}$B$_{2}$ \cite{yamamoto}, MgB$_{2-x}$Be$_{x}$ \cite{ahn}
are additional potential
candidates, as well as overdoped cuprates \cite%
{niedermayer,uemuraod,bernhard}. In particular, evidence for
$1/\lambda \left( 0\right) ^{2}\propto T_{c}^{2}$ emerges for
NbB$_{2+x}$ from the muon-spin rotation study of Khasanov \textit{et
al}. \cite{khasanov}. Moreover, based on our analysis, a plot of
$T_{c}$ \textit{vs}. $1/\lambda \left( 0\right) ^{2}$ of cuprate
superconductors should rise more steeply in the underdoped limit
($T_{c}\propto 1/\lambda \left( 0\right) ^{2}$) than in the
overdoped limit ($T_{c}\propto 1/\lambda \left( 0\right) $). Various
experiments appear to support this behavior qualitatively
\cite{Uemura89,Uemura91,niedermayer,uemuraod,bernhard} but more data
are necessary to confirm it quantitatively. On the other hand, there
is considerable experimental evidence \cite{peets} that for
overdoped cuprates the zero temperature gap is proportional to
$T_{c}$ (Eq.(\ref{eq3})).

\section{Acknowledgments}
The author is grateful to R. Khasanov and H. Keller for very useful
comments and suggestions on the subject matter.

\end{document}